\documentclass[12pt]{article}
\usepackage{amsmath}
\usepackage{amsfonts}
\usepackage{mathptm}
\usepackage{graphicx}
\usepackage{caption}
\newcommand{\beq}{\begin{equation}}
\newcommand{\eeq}{\end{equation}}
\newcommand{\bea}{\begin{eqnarray}}
\newcommand{\eea}{\end{eqnarray}}
\newcommand{\beas}{\begin{eqnarray*}}
\newcommand{\eeas}{\end{eqnarray*}}

\newcommand{\epm}{e^+e^-}

\begin{document}
\thispagestyle{empty}
\voffset -2cm
\begin{flushright}
March 2025\\
Revised:\\
May 2025\\
\vspace*{1.cm}
\end{flushright}
\begin{center}
{\Large\bf Monte Carlo phase space integration of multiparticle cross sections
with {\tt carlomat\_4.5}}\\
\vspace*{1.5cm}
Karol Ko\l odziej\footnote{E-mail: karol.kolodziej@us.edu.pl}\\[1cm]
{\small\it
Institute of Physics, University of Silesia\\ 
ul. 75 Pu\l ku Piechoty 1, PL-41500 Chorz\'ow, Poland}\\
\vspace*{1.5cm}
{\bf Abstract}\\
\end{center}
Multidimensional phase space integrals must be calculated in order to obtain
predictions for total or differential cross sections, or to simulate unweighted 
events of multiparticle reactions. The corresponding matrix elements, already 
in the leading order, receive contributions typically from dozens of thousands 
of the Feynman diagrams, many of which often involve strong peaks due to 
denominators of some Feynman propagators approaching their minima. 
As the number of peaks exceeds by far the number of integration variables, 
such integrals can practically be performed within the multichannel Monte Carlo 
approach, with different phase space parameterizations, each designed
to smooth possibly a few peaks at a time. This obviously requires a lot different
phase space parameterizations which, if possible, should be generated and combined
in a single multichannel Monte Carlo procedure in a fully automatic way.
A few different approaches to the calculation of the multidimensional phase 
space integrals have been incorporated in version 4.5 of the multipurpose
Monte Carlo program {\tt carlomat}. The present work illustrates how 
{\tt carlomat\_4.5} can facilitate the challenging task of calculating
the multidimensional phase space integrals.
\vfill

\newpage

\section{Introduction}
Various aspects of the theory of fundamental interactions, such as 
the non-Abelian nature of gauge symmetry group or
the mechanism of the symmetry breaking can be studied in high energy colliders
through observations of processes of a few heavy particle production at a time 
which, if combined with their almost immediate decays, lead to multiparticle 
final states.
In order to fully investigate the nature and interactions of the heavy particles 
produced, the corresponding multiparticle reactions must be measured, 
including their distributions and spin correlations. 
Such measurements can be best performed in a clean experimental environment of
the planned electron-positron colliders, as
the Future Circular Collider (FCC--ee)~\cite{FCC} and Compact Linear 
Collider (CLIC)~\cite{CLIC} at CERN, the International Linear Collider 
(ILC)~\cite{ILC} in Japan, or the Circular Electron--Positron Collider 
(CEPC)~\cite{CEPC} in China.

Multiparticle reactions must also be measured and compared with model predictions 
at low energy $\epm$ colliders
in order to determine more precisely hadronic contributions to the vacuum 
polarization through dispersion relations from the ratio 
$R=\sigma(\epm\to \text{hadrons})/\sigma(\epm\to \mu^+\mu^-)$ 
at the centre of mass energies below  the $J/\psi$ production threshold.
The hadronic contributions to the vacuum polarization 
have an impact on precision of theoretical predictions for the muon $g-2$ 
anomaly and play an important role in the evolution of 
the fine structure constant $\alpha(Q^2)$ from the Thomson limit to high energy 
scales. Precision knowledge of $\alpha(m_Z^2)$ would be vital for the Giga Z option
of any future $\epm$ collider.

Derivation of theoretical predictions for cross sections or asymmetries
of any multiparticle reaction requires integration  over a 
multidimensional phase space of a squared modulus of the corresponding 
matrix element, which often receives contributions from several dozens of 
thousands or even several hundreds of thousands of the Feynman diagrams. 
Such multidimensional integrals can be in practice calculated
only with the Monte Carlo (MC) method. Whenever denominators of the Feynman 
propagators approach its minimum, the corresponding amplitudes may become strongly 
peaked. In order to obtain reliable results of the integration,
those peaks must be smoothed by appropriate changes of the integration
variables. However, the number of peaks in the full amplitude of the reaction
usually substantially exceeds the number of variables in the corresponding 
differential phase space element parameterization. Therefore, 
the  multichannel MC approach must be used, where the name {\em channel} 
refers to a single phase space parameterization which can smooth possibly a few
peaks at a time. All different parameterizations must be then combined into
a single parameterization that is used in the MC integration. As the number
of channels is typically very large, the whole process of generating appropriate
multichannel differential phase space parameterization must be fully automized.

An obvious way to follow in order to map out all the peaks is to generate one
subroutine containing the
phase space parameterization for each individual Feynman diagram, as it 
was originally done in {\tt carlomat\_1.0} \cite{carlomat}. 
However, for multiparticle reactions, this approach leads to a large number of 
subroutines containing different parameterizations and the resulting multichannel 
phase space routine is huge indeed and usually difficult to compile. Needless to say 
that the execution time of the MC integration would also become rather long.
A modification of this approach was introduced in {\tt carlomat\_2.0} 
\cite{carlomat2}, where several phase space parameterizations corresponding to
the Feynman diagrams of the same topology were combined into a single subroutine
which resulted in a substantially shorter multichannel MC integration
routine. The efficiency of that approach was further improved in 
{\tt carlomat\_4.0} \cite{carlomat4} by automatic inclusion of parameterizations 
which map away the t-channel poles and peaks due to soft and collinear photon or 
gluon emission. However, for some multiparticle reactions as, e.g., $2\to 8$ particle
scattering which are
relevant for the associated production of the top quark pair and the Higgs or
vector boson, the resulting multichannel MC kinematics routine may be still 
difficult to compile and would need quite a long execution time.
To overcome these difficulties a different approach was proposed 
in {\tt PSGen} \cite{PSGen}, a program for generation of phase space 
parameterizations for the multichannel MC integration, where the phase 
space parameterizations 
of a given reaction are generated automatically according to predefined patterns
which are supposed to smooth only the most relevant peaks of the matrix element. 
This reduces substantially the size of the multichannel MC kinematics routine 
which can be very fast generated and compiled and executed in a much shorter time. 
However, it is obvious that, as not all the peaks present in the matrix element
are taken into account by {\tt PSGen}, some loss of the MC integration 
convergence should be expected. 

In order to facilitate the challenging task of calculating
the multidimensional phase space integrals, {\tt carlomat\_4.5}, a new version 
of the multipurpose Monte Carlo program {\tt carlomat} has been written.  
It allows to calculate the cross section either with the kinematics routine
generated by {\tt carlomat}, or with the kinematics routine generated by 
{\tt PSGen\_1.1}, the current version of {\tt PSGen}, dependent on user's choice.
The kinematics chosen can be automatcally combined with the leading order (LO) 
matrix element generated by {\tt carlomat} or with the user provided matrix element, 
either in the LO or in higher orders. The MC integration can be performed 
either with {\tt carlos}, a plain MC integration routine of
{\tt carlomat} \cite{carlomat}, or  {\tt VEGAS} \cite{VEGAS} as the latter has been
implemented in {\tt carlomat\_4.5}. {\tt VEGAS} handles peaks of the integrand with 
an importance sampling technique which is based on appropriate adaptation of 
the integration grid in subsequent iterations of the integral.
As the original version of {\tt VEGAS} \cite{VEGAS} is limited to calculation
of integrals up to 10 dimensions, its Fortran source has been modified by 
the author of the present work so that it can also be used to calculate 
integrals of higher dimension. However, as it will be discussed later on in 
Section~3, its use may then encounter some problems.

In the present work, a few issues concerning efficiency and convergence of the
MC integration will be addressed by comparing results for the
cross sections of a few physically
interesting multiparticle reactions that could be measured at any future high 
energy $\epm$ collider. The cross sections will be calculated with different 
phase space parameterizations generated automatically with the above described 
algorithms.
It will also be checked to which extent different options of performing the actual 
MC integration, such as the initial scan of the generated kinematics channels or
an adaptation of integration weights after each iteration of the integral, 
or the use of adaptive MC integration routine {\tt VEGAS}
influence the integration efficiency.

The article is organized as follows. Some calculational details and
useful hints concerning usage of {\tt carlomat\_4.5} are given in 
Section~2. Section~3 contains a sample of cross sections which should illustrate 
possible problems related to the calculation of multidimensional MC integrals.
The conclusions are formulated in Section~4.

\section{Calculational details and program usage}
In this section, some details on generation of the code, preparation for running
the MC program and selection of options for numerical calculation
of the cross sections presented in Section~3 are given. 

The user defines the reaction to be considered and chooses the way in which 
the phase space parameterizations 
should be generated by an appropriate choice of flag {\tt ipsgen} 
in {\tt carlomat.f}, the main program of the code generation package of
{\tt carlomat\_4.5} \cite{carlomat4.5}, the current version of 
 {\tt carlomat}. If integer variable {\tt ipsgen} is set to 
any value different from 1 then the kinematics routine will be
automatically generated by {\tt carlomat}, else, if {\tt ipsgen=1}, then
the kinematics routine should be generated by {\tt PSGen}, in the way described 
below. 
There are a few other flags in {\tt carlomat.f} that should be set to desired
values and then the program should be run with the command\\
{\tt make code}.\\
Note that the Fortran compiler to be used is chosen in a corresponding 
{\tt makefile}. 
If {\tt ipsgen=1} then the user should
switch to {\tt PSGen/code\_generation} and run the phase generation 
program there, again with the command\\
{\tt make code}.\\
Prior to it, some flags described in the main program {\tt PSGen} can be 
selected and the phase generation patterns can be edited in file 
{\tt genps.dat} in order to map peaks of the considered reaction in the best way.
How those patterns are to be defined is described in detail in \cite{PSGen}.
Note that any predefined pattern can be commented out by
setting the first integer entry of the corresponding line to 0. If the width of 
a massive 
particle is set to a character variable {\tt zero}, then the corresponding squared
four momentum transfer in the Feynman propagator
will be generated according to a flat probability
distribution. 

The MC integration is performed with the automatically generated probability 
density function $f(x)$ which is defined in terms of probability density 
functions $f_i(x)$, $i=1,...,n_{\text{kin}}$, also automatically generated, 
in the following way 
\bea
\label{distrib}
f(x)=\sum_{i=1}^{n_{\text{kin}}}a_if_i(x), 
\eea
where $x=(x_1,...,x_{n_{\text{d}}})$, $0<x_i<1$, are random numbers and weights 
$a_i\ge 0$, $i=1,...,n_{\text{kin}}$, satisfy the condition 
$\sum\limits_{i=1}^{n_{\text{kin}}}a_i=1$. Densities $f(x)$ and $f_i(x)$ of
Eq.~(\ref{distrib}) must fulfil the following normalization conditions 
\bea
\label{normal}
\int\limits_0^1\text{d}^{n_{\text{d}}}f(x)=\int\limits_0^1\text{d}^{n_{\text{d}}}f_i(x)=
\text{vol(Lips)},
\eea
where vol(Lips) is the total volume of the Lorentz invariant phase space of 
the considered reaction. 
Parameterizations of differential phase space elements $f_i(x)$ of 
Eq.~(\ref{distrib}) are generated either with 
{\tt carlomat\_4.5} \cite{carlomat4.5}
or with {\tt PSGen\_1.1} \cite{PSGen1.1}, the current version of
{\tt PSGen}.
The actual probability density function $f_i(x)$ according to which the final 
state particle four momenta are generated, which are needed to calculate the 
corresponding matrix element or to be stored as MC events, 
is chosen from the set
$\{f_j(x),\;\; j=1,...,n_{\text{kin}}\}$ if uniformly
distributed random number $\xi\in[0,1]$ falls into the interval
$a_0+...+a_{i-1} \leq \xi \leq a_0+...+a_{i}$, with $a_0=0$.
In the present work, the corresponding LO standard model (SM) matrix element 
is generated by {\tt carlomat\_4.5}.

Once the code for calculation of the matrix element and the kinematics routines
have been generated, the user should choose the centre of mass
energies and set the desired options, by appropriately editing the main MC
program {\tt carlocom\_mpi.f} in directory {\tt carlomat\_4.5/mc\_computation}. 
Then the program can be run with the command\\
{\tt make -f mpi mc}.\\ 
The output will be written to files {\tt tot\_i\_...}, where 
{\tt i=0,1,2,...,n\_proc} 
labels computational processes within the Message Passing Interface (MPI)
whose number {\tt n\_proc} should be set in the first line of the makefile 
{\tt mpi}.

A number of options are available in the main program {\tt carlocom\_mpi.f}
which allow to better control the MC integration. One of them is governed by
flag {\tt iscan}. If {\tt iscan=1} then the MC integral is scanned. 
This means that prior to the actual calculation with a large number of calls 
to the integrand and, e.g. 10 iterations, it is calculated in one iteration with 
a relatively small, say 1000, number of calls, each time with a single phase 
space parameterization $f_i(x)$. The latter is selected by setting $a_i=1$ 
and all other weights, $j\neq i$, $a_j=0$. Denote the
rough estimate of the cross section obtained in this way by $\sigma_i$. Then
weights $a_i$ of Eq.~(\ref{distrib}) for the first iteration of the MC integral 
are determined according to the following formula
\bea
\label{weights}
a_i=\frac{\sigma_i}{\sum\limits_{i=1}^{n_{\text{kin}}}\sigma_i}, \qquad
i=1,...,n_{\text{kin}}.
\eea
On the other hand, if {\tt iscan=0} then the first iteration of the MC integral
is calculated with equal weights, $a_i=1/n_{\text{kin}}$. Another flag in
{\tt carlocom\_mpi.f} is 
{\tt iwadapt}. If {\tt iwadapt=1} then the weights $a_i$ are calculated
anew after each iteration according to Eq.~(\ref{weights}), with $\sigma_i$ 
being a collection of all contributions to the total cross section
obtained if probability density function $f_i(x)$ has been selected for 
calculation of the final state particle momenta. Else, if {\tt iwadapt=0} then
all the iterations of the MC integral are calculated with the weights $a_i$
fixed at the very beginning, i.e. before the first iteration of the MC integral.
\section{Some illustrative results}
In this section, the efficiency of different approaches to calculation
of multidimensional integrals with the MC method is examined. As illustrative 
examples, the LO SM cross sections of a few physically 
interesting multiparticle reactions, which could potentially be measured at any 
future high energy $\epm$ collider, are considered.
In particular, cross sections of the following reactions
\begin{align}
\label{WW}
\epm &\to\; \mu^+\nu_{\mu}\mu^-\bar{\nu}_{\mu}, & n_{\text{d}}&=8,
&\text{19 diagrams,}&\\  
\label{tt}
\epm &\to \; b\mu^+\nu_{\mu}\bar{b}\mu^-\bar{\nu}_{\mu}, & n_{\text{d}}&=14,
& \text{452 diagrams},&\\ 
\label{ttH}
\epm &\to \; b \bar{b} b\mu^+\nu_{\mu}\bar{b}\mu^-\bar{\nu}_{\mu}, 
& n_{\text{d}}&=20, &\text{46890 diagrams},&
\end{align}
where dimension $n_{\text{d}}$ of the corresponding phase space integral and 
the number of the LO SM Feynman diagrams are indicated on the right hand side 
of each reaction, are calculated. The final states of reactions (\ref{WW}), (\ref{tt}) 
and (\ref{ttH}) represent relatively clean detection channels of, respectively, 
$W^+W^-$, top quark pair production and associated production of the Higgs 
boson and top quark pair. To enable their identification 
the following cuts:
\bea 
\label{cuts}
\begin{split}
5^{\circ}<\theta(\text{l, beam}),\;\theta(\text{q, beam})< 175^{\circ}, 
&\quad& \theta(\text{l, l'}),\;\theta(\text{q, q'}),
\;\theta(\text{q, l})> 10^{\circ}, \\
E_{\text{l}},\;E_{\text{q}}> 15\,\text{GeV}, &\quad& 
E_{T{\text{missing}}} > 15\,\text{GeV},\qquad\qquad\;\;
\end{split}
\eea
where $\text{l},\;\text{l}'$ stand for either $\mu^-$ or $\mu^+$ and 
$\text{q},\;\text{q}'$ stand for either $b$ or $\bar{b}$, are imposed.

In order to find out the optimal probability density function of 
Eq.~(\ref{distrib}), the proper choice of flags and the adequate MC 
integration routine, the results for the LO cross sections 
of reactions (\ref{WW}) and (\ref{tt}) at 
$\sqrt{s}=360\,\text{GeV},500\,\text{GeV}$, $800\,\text{GeV and }1\,\text{TeV}$ 
are collected in tables \ref{tab:WW} and \ref{tab:tt}, respectively. 
In both tables, all entries in columns 3--6 and rows with the same value of 
$\sqrt{s}$ in column 1 show the cross sections calculated with 
different phase space parameterizations and various choices of options for the MC 
integration, as described in the following.
Values of {\tt ivegas/ipsgen} listed column 2 correspond to
choices of the flags described in Section~2. In particular, if the first
integer in column 2, i.e. {\tt ivegas}, is equal to {\tt 0(1)} then the 
integration is performed with a plane MC integration routine {\tt carlos} 
(an adaptive MC integration routine {\tt VEGAS}).
The other integer in column 2 indicates whether the multichannel probability
density function $f(x)$ of Eq.~(\ref{distrib}) has been generated by
{\tt carlomat\_4.5} ({\tt ipsgen=0}) or by {\tt PSGen\_1.1} ({\tt ipsgen=1}). 
The two upper rows of columns 3--6 specify choices of flags {\tt iscan} and 
{\tt iwadapt}, described in Section~2, which have been used in the MC integration
of cross sections listed below.


{\small
\begin{table}[ht]
\begin{center}
\begin{tabular}{cccccc}
\hline
\rule{0mm}{5mm}$\sqrt{s}$ &{\tt ivegas/}  & {\tt iscan=0} &{\tt iscan=1} & {\tt iscan=0} 
& {\tt iscan=1}\\
(GeV)   & {\tt ipsgen}& {\tt iwadapt=0} &{\tt iwadapt=0} & {\tt iwadapt=1} & 
{\tt iwadapt=1}\\[2mm]
\hline
\rule{0mm}{5mm} 360 & {\tt 0/0}& 111.06(45) & 111.48(17) & 111.16(16) & 111.74(15) \\
 360 & {\tt 1/0}& 106.10(18) & 111.69(5)  & 111.61(4)  & 111.62(4)  \\
 360 & {\tt 0/1}& 111.58(42) & 111.71(16) & 111.65(16) & 111.53(15) \\
 360 & {\tt 1/1}& 119.66(18) & 112.57(5)  & 111.88(4)  & 111.81(4)  \\[2mm]
 500 & {\tt 0/0}& 70.51(44)  & 70.85(16)  & 70.42(15)  & 70.55(15) \\
 500 & {\tt 1/0}& 66.87(13)  & 70.62(4)   & 70.48(3)   & 70.50(3)  \\
 500 & {\tt 0/1}& 70.18(40)  & 70.66(15)  & 70.67(15)  & 70.36(14) \\
 500 & {\tt 1/1}& 74.79(13)  & 70.98(4)   & 70.59(3)   & 70.58(3)  \\[2mm]
 800 & {\tt 0/0}& 31.95(31)  & 32.30(11)  & 32.25(11) & 32.11(11) \\
 800 & {\tt 1/0}& 30.38(6)   & 32.23(2)   & 32.19(2)  & 32.19(2)  \\
 800 & {\tt 0/1}& 31.93(28)  & 32.20(11)  & 32.23(11) & 32.08(10) \\
 800 & {\tt 1/1}& 33.24(6)   & 32.22(2)   & 32.25(2)  & 32.26(2)  \\[2mm]
 1000$\;\;$ & {\tt 0/0}& 21.09(24) & 20.90(8)   & 20.93(8) & 20.89(8) \\
 1000$\;\;$  & {\tt 1/0}& 19.66(4)  & 20.88(1)   & 20.89(1) & 20.90(1) \\
 1000$\;\;$ & {\tt 0/1}& 20.75(21) & 20.87(8)   & 20.95(8) & 20.93(8) \\
 1000$\;\;$  & {\tt 1/1}& 21.59(4)  & 20.82(1)   & 20.92(1) & 20.92(1) \\[2mm]
\hline
\end{tabular}
\caption{LO cross sections in fb of reaction (\ref{WW}) calculated with different 
choices of options for the MC integration, as described in the main text. Uncertainties 
of the last digits are given in parentheses.\label{tab:WW}}
\end{center}
\end{table}}

A brief inspection of table~\ref{tab:WW} shows that the initial scan of the
generated kinematic channels reduces the standard deviation of the MC integral
by roughly a factor 3. The same observations holds also for the use of adaptive
MC integration routine {\tt VEGAS}. If, in addition, the weight adaptation
is turned on, then the MC error is further reduced, but to much less extent. 
Note, however, that the combination
{\tt ivegas/ipsgen=1/0} ({\tt ivegas/ipsgen=1/1}) with {\tt iscan=0} and 
{\tt iwadapt=0} gives a small error with an underestimation (overestimation) 
of the integral, which is not compatible with the other results. This is most
probably because the {\tt VEGAS} grid adaptation algorithm is accidentally 
caught in some kinematics channels, with practically  no possibility of 
choosing the other channels in consecutive iterations. 

\begin{figure}[ht]
\vspace*{-6cm}
\begin{tabular}{cc}
 \hspace*{-3.5cm}\includegraphics[width=15cm]{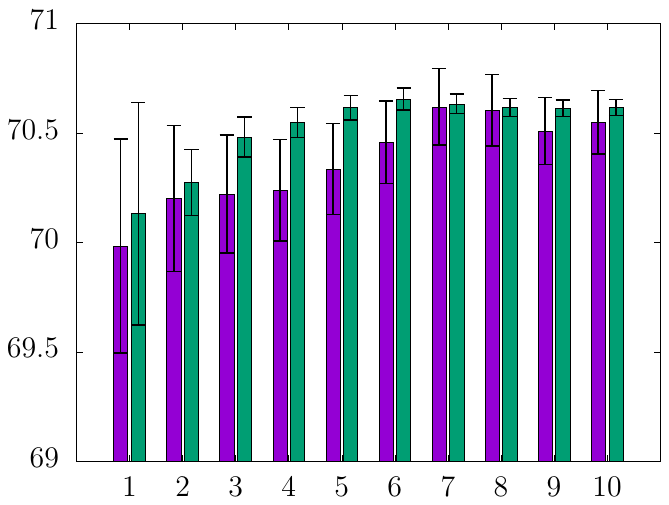}&
 \hspace*{-6.5cm}\includegraphics[width=15cm]{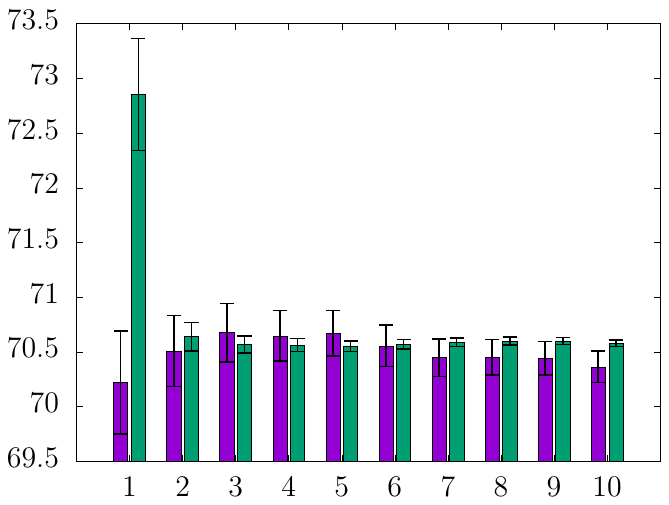}
\end{tabular}
\vspace*{-8cm}
\caption{Accumulated results for the LO SM cross section in fb of reaction (\ref{WW}) at 
$\sqrt{s}=500$~GeV as functions of the number of 
iterations. The results plotted in the left panel have been integrated with 
$f(x)$ of {\tt carlomat\_4.5} and those plotted in the 
right panel with $f(x)$ of {\tt PSGen\_1.1}. In both panels, the left histogram has 
been integrated with
{\tt carlos} and the right histogram with {\tt VEGAS}, with {\tt iscan=1} 
and {\tt iwadapt=1}}
\label{fig:WW}
\end{figure}

Accumulated results for the LO SM cross section of reaction (\ref{WW}) at 
$\sqrt{s}=500$~GeV as functions of the number of 
iterations are shown in Fig.~\ref{fig:WW}. The results plotted in the left panel 
have been integrated with the probability density function $f(x)$ of Eq.~(\ref{distrib}) 
generated by {\tt carlomat\_4.5} 
and those plotted in the right panel with $f(x)$ generated by {\tt PSGen\_1.1}. In both
panels, the left histogram shows
the results integrated with {\tt carlos} while the right histogram depicts the
results obtained with {\tt VEGAS} and {\tt iscan=1} and {\tt iwadapt=1} have been assumed. 
By comparing the left and right histograms
in both panels of Fig.~\ref{fig:WW}, we see that the {\tt VEGAS} algorithm  
reduces the standard deviation in consecutive iterations much better than that 
of {\tt carlos}. One should also note that the result of the first iteration in
the right panel departs substantially from the results of further iterations.
This is because of the fact that the $f(x)$ generated
by {\tt PSGen\_1.1} does not contain mappings of all the peaks of the
integrand. 

\begin{table}
\begin{center}
\begin{tabular}{cccccc}
\hline
\rule{0mm}{5mm}$\sqrt{s}$ &{\tt ivegas/}  & {\tt iscan=0} &{\tt iscan=1} & {\tt iscan=0} & {\tt iscan=1}\\
(GeV)   & {\tt ipsgen}& {\tt iwadapt=0} &{\tt iwadapt=0} & {\tt iwadapt=1} & 
{\tt iwadapt=1}\\[2mm]
\hline
\rule{0mm}{5mm} 360 & {\tt 0/0}& 4.3212(222) & 4.3154(25) & 4.3224(31) & 4.3148(23) \\
 360 & {\tt 1/0}& 4.2001(215) & 4.2509(33)  & 4.2989(22)  & 4.2991(29)  \\
 360 & {\tt 0/1}& 4.3281(123) & 4.3216(25) & 4.3124(17) & 4.3157(19) \\
 360 & {\tt 1/1}& 4.3497(110) & 4.4967(18)  & 4.3337(17)  & 4.3234(13)  \\[2mm]
 500 & {\tt 0/0}& 5.7721(334)  & 5.7444(45)  & 5.7584(52)  & 5.7416(45) \\
 500 & {\tt 1/0}& 5.3242(326)  & 6.2811(23)   & 5.7385(70)   & 5.7384(35)  \\
 500 & {\tt 0/1}& 5.7628(173)  & 5.7625(28)  & 5.7606(30)  & 5.7618(26) \\
 500 & {\tt 1/1}& 6.0091(155)  & 5.8128(23)   & 5.7627(25)   & 5.7644(22)  \\[2mm]
 800 & {\tt 0/0}& 2.8451(214)  & 2.8395(56)  & 2.8585(93) & 2.8420(68) \\
 800 & {\tt 1/0}& 2.4527(352)   & 3.2007(20)   & 2.8013(44)  & 2.7906(58)  \\
 800 & {\tt 0/1}& 2.8583(91)  & 2.8662(20)  & 2.8647(22) & 2.8688(20) \\
 800 & {\tt 1/1}& 3.0329(83)   & 3.0706(12)   & 2.8634(17)  & 2.8625(15)  \\[2mm]
 1000$\;\;$ & {\tt 0/0}& 1.9306(202) & 1.9433(68)   & 1.9363(77) & 1.9230(69) \\
 1000$\;\;$  & {\tt 1/0}& 2.3477(276)  & 2.0881(9)   & 0.4127(430) & 1.9841(50) \\
 1000$\;\;$ & {\tt 0/1}& 1.9675(65) & 1.9644(18)   & 1.9634(18) & 1.9621(17) \\
 1000$\;\;$  & {\tt 1/1}& 2.0288(60)  & 2.0864(6)   & 1.8903(42) & 1.8667(20) \\[2mm]
\hline
\end{tabular}
\caption{LO cross sections in fb of (\ref{tt}) calculated with different choices 
of options for the MC integration, as described in the main text. Uncertainties 
of the last digits are given in parentheses.\label{tab:tt}}
\end{center}
\end{table}

Looking at table~\ref{tab:tt}, one sees that the initial scan of the
generated kinematic channels reduces the standard deviation of the MC integral
even more substantially than in table~\ref{tab:WW}. However, the error reduction 
due to the use of 
adaptive integration routine {\tt VEGAS} is not as illuminating as in 
table~\ref{tab:WW}. In contrary, the results for $\sqrt{s}= 800\,\text{GeV}$ and
$\sqrt{s}=1\,\text{TeV}$  obtained with the use of {\tt VEGAS} do not seem to 
be reliable. It looks as if the {\tt VEGAS} grid adaptation 
algorithm does not work as efficiently for $n_{\text{d}}=14$,
as it does for $n_{\text{d}}\leq 10$, which it was originally designed for.
This conjecture seems to be confirmed for reaction (\ref{ttH}), the cross sections 
of which integrated over the 20-dimensional phase space with {\tt VEGAS}
are not reliable at all. However, it is also possible that the {\tt VEGAS}
grid adaptation algorithm does not conform well with large number of kinematics 
channels for reactions (\ref{tt}) and (\ref{ttH}) which are selected randomly 
during the computation of integral. This 
has been confirmed by counting calls to different kinematics channels in each
iteration.
After a few iterations {\tt VEGAS} keeps calling the same kinematics all the time
and hence its weight approaches 1. The plain MC integration routine {\tt carlos}
seem to cope better with these problems.

\begin{figure}
\vspace*{-6cm}
\begin{tabular}{cc}
 \hspace*{-3.5cm}\includegraphics[width=15cm]{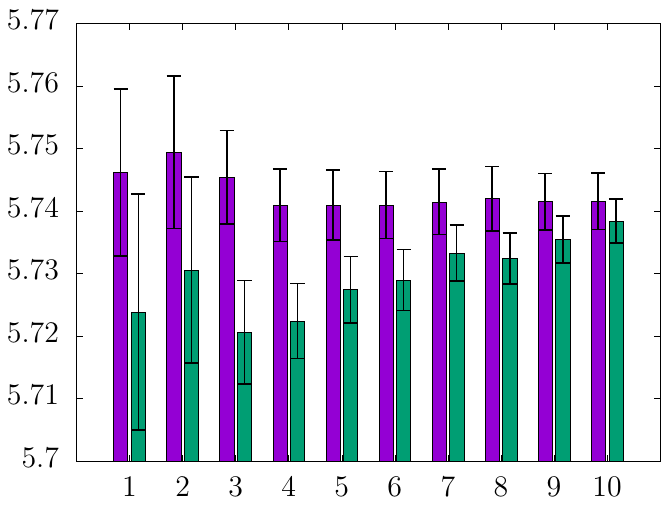}&
 \hspace*{-6.5cm}\includegraphics[width=15cm]{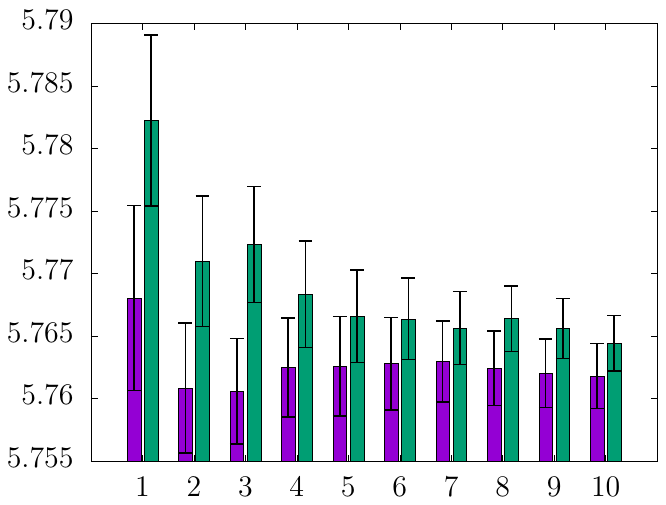}
\end{tabular}
\vspace*{-8cm}
\caption{Accumulated results for the LO SM cross section in fb of reaction (\ref{tt}) at 
$\sqrt{s}=500$~GeV as functions of the number of 
iterations. The results plotted in the left panel have been integrated with 
the phase space generated by {\tt carlomat\_4.5} and those plotted in the 
right panel with {\tt PSGen\_1.1}. In both panels, the left histogram has 
been integrated with
{\tt carlos} and the right histogram with {\tt VEGAS}, with {\tt iscan=1} 
and {\tt iwadapt=1}}
\label{fig:tt}
\end{figure}

Accumulated results for the LO SM cross section of reaction (\ref{tt}) at 
$\sqrt{s}=500$~GeV as functions of the number of iterations 
are shown in Fig.~\ref{fig:tt}. The results plotted in the left panel 
have been integrated with the probability density function $f(x)$ generated by
{\tt carlomat\_4.5} and those plotted in the right panel with $f(x)$ generated by
 {\tt PSGen\_1.1}, in both cases with {\tt iscan=1} and {\tt iwadapt=1}. Again, 
in both panels, the left histogram shows the results integrated with {\tt carlos} 
while the right one depicts the results obtained with {\tt VEGAS}. The advantage 
of the adaptive algorithm of {\tt VEGAS} over the plain MC sampling of {\tt carlos}
in reducing the standard deviation  is not any more as pronounced as in 
Fig.~\ref{fig:WW}, as expected.

\begin{figure}
\vspace*{-6cm}
\begin{tabular}{cc}
 \hspace*{-3.5cm}\includegraphics[width=15cm]{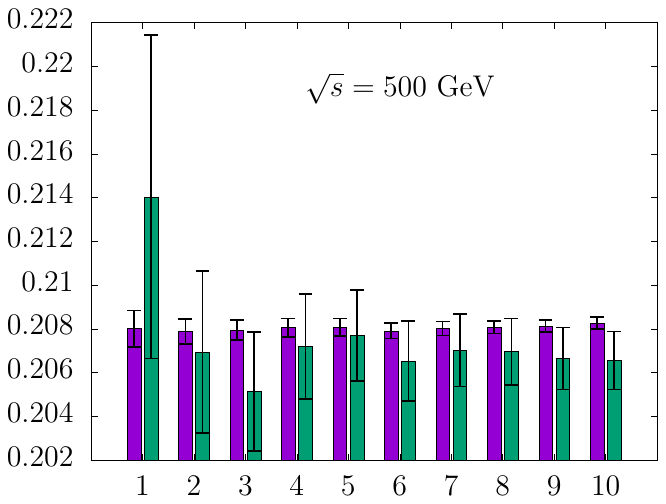}&
 \hspace*{-6.5cm}\includegraphics[width=15cm]{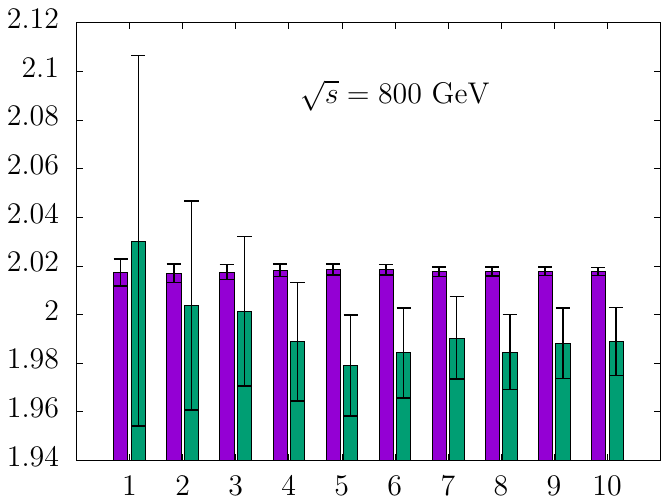}
\end{tabular}
\vspace*{-8cm}
\caption{Accumulated results for the LO SM cross section in fb of reaction (\ref{ttH}) 
at $\sqrt{s}=500$~GeV and $\sqrt{s}=800$~GeV as functions of the number of 
iterations. The left (right) histograms in both panels have been integrated 
with {\tt carlos}, using {\tt iscan=1} and {\tt iwadapt=1}, and the phase space 
generated by {\tt carlomat\_4.5} ({\tt PSGen\_1.1}).}
\label{fig:ttH}
\end{figure}

Therefore, in Fig.~\ref{fig:ttH}, only the results 
obtained with {\tt carlos}, using {\tt iscan=1} and {\tt iwadapt=1}, are shown.
The accumulated results for the cross section of reaction (\ref{ttH}) as functions 
of the number of iterations at $\sqrt{s}=500$~GeV and $\sqrt{s}=800$~GeV are plotted 
in the left and right panel, respectively. In both panels, 
the left (right) 
histograms show results integrated with the probability density function $f(x)$
generated by {\tt carlomat\_4.5} ({\tt PSGen\_1.1}). The advantage of the $f(x)$
of {\tt carlomat\_4.5}, which maps out all peaks of
the integrand, is clearly visible. The results of consecutive iterations of 
the MC integral are
much more stable and the standard deviation is much smaller than in case
of the integration with the $f(x)$ of {\tt PSGen\_1.1} which covers
only the most dominant peaks of the associated top quark pair and Higgs boson
production.

\section{Conclusions}
It has been shown that calculation of multidimensional phase space integrals, 
which are necessary in order to obtain predictions for total or differential cross 
sections, or to simulate unweighted events of different
physically interesting reactions, is a challenging task. It can be in practice solved
only with the Monte Carlo method. 
As the corresponding matrix elements
involve many peaks, the variance of the MC integral can be reduced only if
those peaks are mapped out which is achieved by the use
of the multichannel MC approach, with different phase space parameterizations generated
and combined in the single probability distribution in a fully automatic way.
A few different approaches to this task have been applied in the present work.
It has been shown that there is no single golden recipe to obtain reliable results
for the MC integrals of interest. Which particular approach should be used depends
mostly on the dimension of the phase space integral, but also on the centre of 
mass energy of the considered reaction.

The Fortran code with which the results shown in the present work were obtained 
is public. It can be downloaded from the web pages whose addresses are given
in \cite{carlomat4.5}, \cite{PSGen1.1} and freely used.



\end{document}